\documentclass[preprint,final,5p,times,twocolumn]{elsarticle}
\usepackage{rotating,color,subfigure,amssymb}
\usepackage{alphalph}
\usepackage{amsmath}
\usepackage[T1]{fontenc}
\usepackage{amssymb}
\journal{Physics Letters B}
\begin{document}
                                       
\begin{frontmatter}

\title{Neutron polarisation transfer, $C_{x'}^n$, in $\pi^+$ photoproduction off the proton}
\date{\today}

\author[UoY]{M.~Bashkanov}\ead{mikhail.bashkanov@york.ac.uk}
\author[UoY]{D.P.~Watts}

\author[UoR]{S.J.D.~Kay}

\author[UBasel]{S.~Abt}
\author[KPM]{P.~Achenbach}
\author[KPM]{P.~Adlarson}
\author[UoB]{F.~Afzal}
\author[UoR]{Z.~Ahmed}
\author[KSU]{C.S.~Akondi}
\author[UoG]{J.R.M.~Annand}
%
\author[UoB]{R.~Beck}
\author[KPM]{M.~Biroth}
\author[Dubna]{N.~Borisov}
\author[INFN]{A.~Braghieri}
\author[GWU]{W.J.~Briscoe}
%
\author[KPM]{F.~Cividini}
\author[SMU]{C.~Collicott}
\author[Pavia,INFN]{S.~Costanza}
%
\author[KPM]{A.~Denig}
\author[GWU]{E.J.~Downie}
\author[Giessen,KPM]{P.~Drexler}

\author[UoY]{S.~Fegan}

\author[Tomsk]{A.~Fix}

%
%
\author[UoG]{S.~Gardner}
\author[UBasel]{D.~Ghosal}
\author[UoG]{D.I.~Glazier}
\author[Dubna]{I.~Gorodnov}
\author[KPM]{W.~Gradl}
\author[INR]{D.~Gurevich}
%
\author[KPM]{L. Heijkenskj{\"o}ld}
\author[MAU]{D.~Hornidge}
\author[UoR]{G.M.~Huber}
%
\author[KPM,Dubna]{V.L.~Kashevarov}

\author[RBI]{M.~Korolija}
\author[UBasel]{B.~Krusche}
%

\author[Dubna]{A.~Lazarev}
\author[UoG]{K.~Livingston}
\author[UBasel]{S.~Lutterer}
%
\author[UoG]{I.J.D.~MacGregor}
\author[KSU]{D.M.~Manley}
\author[KPM,MAU]{P.P.~Martel}
\author[UoM]{R.~Miskimen}
\author[KPM]{E.~Mornacchi}
%
\author[UoG]{C. Mullen}
\author[Dubna]{A.~Neganov}
\author[KPM]{A.~Neiser}
%
\author[KPM]{M.~Ostrick}
\author[KPM]{P.B.~Otte}
%
\author[UoR]{D.~Paudyal}
\author[INFN]{P.~Pedroni}

%
\author[Basel]{T.~Rostomyan}
%
%
\author[KPM]{V.~Sokhoyan}
\author[UoB]{K.~Spieker}
\author[KPM]{O.~Steffen}
\author[GWU]{I.I.~Strakovsky}
\author[UBasel]{T.~Strub}
\author[RBI]{I.~Supek}
%
\author[UoB]{A.~Thiel}
\author[KPM]{M.~Thiel}
\author[KPM]{A.~Thomas}
%
\author[Dubna]{Yu.A.~Usov}
%
\author[KPM]{S.~Wagner}
\author[UBasel]{N.K.~Walford}
\author[KPM]{J.~Wettig}
\author[KPM]{M.~Wolfes}
%
\author[UoY]{N.~Zachariou}

\address[UoY]{Department of Physics, University of York, Heslington, York, Y010 5DD, UK}
\address[UoR]{University of Regina, Regina, SK S4S-0A2 Canada}
\address[KSU]{Kent State University, Kent, Ohio 44242, USA}
\address[UoG]{SUPA School of Physics and Astronomy, University of Glasgow, Glasgow, G12 8QQ, UK}
\address[UBasel]{Department of Physics, University of Basel, Ch-4056 Basel, Switzerland}
\address[KPM]{Institut f\"ur Kernphysik, University of Mainz, D-55099 Mainz, Germany}
\address[UoB]{Helmholtz-Institut f\"ur Strahlen- und Kernphysik, University Bonn, D-53115 Bonn, Germany}

\address[Dubna]{Joint Institute for Nuclear Research, 141980 Dubna, Russia}

\address[INFN]{INFN Sezione di Pavia, I-27100 Pavia, Pavia, Italy}

\address[GWU]{Center for Nuclear Studies, The George Washington University, Washington, DC 20052, USA}
\address[SMU]{Department of Astronomy and Physics, Saint Mary's University, E4L1E6 Halifax, Canada}
\address[Pavia]{Dipartimento di Fisica, Universit\`a di Pavia, I-27100 Pavia, Italy}
\address[Giessen]{II. Physikalisches Institut, University of Giessen, D-35392 Giessen, Germany}
\address[INR]{Institute for Nuclear Research, RU-125047 Moscow, Russia}

\address[MAU]{Mount Allison University, Sackville, New Brunswick E4L1E6, Canada}
\address[RBI]{Rudjer Boskovic Institute, HR-10000 Zagreb, Croatia}

\address[UoM]{University of Massachusetts, Amherst, Massachusetts 01003, USA}
\address[UoC]{University of California Los Angeles, Los Angeles, California 90095-1547, USA}
\address[RIP]{Racah Institute of Physics, Hebrew University of Jerusalem, Jerusalem 91904, Israel}
\address[RU]{Department of Physics and Astronomy, Rutgers University,
Piscataway, New Jersey, 08854-8019}

\address[JLab]{Jefferson Lab, 12000 Jefferson Ave., Newport News, VA 23606, USA}

\address[Tomsk]{Tomsk Polytechnic University, 634034 Tomsk, Russia}


\cortext[coau]{Corresponding author }

\begin{abstract}
We report a first measurement of the double-polarisation observable, $C_{x'}$, in $\pi^+$ photoproduction off the proton. The $C_{x'}$ double-polarisation observable represents the transfer of polarisation from a circularly polarised photon beam to the recoiling neutron. The MAMI circularly polarised photon beam impinged on a liquid deuterium target cell, with reaction products detected in the Crystal Ball calorimeter. Ancillary apparatus surrounding the target provided tracking, particle identification and determination of recoil nucleon polarisation. The $C_{x'}$ observable is determined for photon energies 800-1400 MeV, providing new constraints on models aiming to elucidate the spectrum and properties of nucleon resonances. This is the first determination of any polarisation observable from the beam-recoil group of observables for this reaction, providing a valuable constraint and systematic check of the current solutions of partial wave analysis based theoretical models.

\end{abstract}

\end{frontmatter}


\section{\label{sec:Int}Introduction}

Photoinduced reactions on proton and neutron targets have played a key role in progressing our knowledge of the excited nucleon spectrum the past decade~\cite{PDG2020}, catalysed by quality nucleon photoproduction data from MAMI, JLAB, ELSA and other facilities~\cite{Thiel}. These have provided a step change in the number of measured observables, statistical accuracy, and kinematic coverage. 

Pion photoproduction is the simplest photoinduced reaction on the nucleon. The reaction can be described theoretically with four complex amplitudes, which can be fully constrained by kinematically complete measurements of a chosen set of eight observables taken from the cross section, single polarisation observables (where the polarisation of either photon beam, target or recoiling nucleon is determined) and double-polarisation observables formed from simultaneous determination of two of the above polarisation quantities. Recent work has indicated that the properties of the different partial waves in the reaction may converge with fewer measurements than the mathematically complete eight, as discussed in Ref.~\cite{BeckPol}. 

Previous double-polarisation measurements for $n\pi^+$ photoproduction are limited to the beam-target group of observables. Measurements of the $G$ observable (linearly polarised beam and transversely polarised target)~\cite{ZA19}, the $E$ observable (circularly polarised beam and longitudinally polarised target~\cite{EClas}) and more limited data sets for $H$ (circularly polarised beam and transversely polarised target) have recently been obtained\cite{BoGa_DB,SAID}. These double-polarisation data, combined with the cross section and single polarisation observables ($\Sigma$, $T$ and $P$)~\cite{ZA19,SAID_MA19} comprise the current world data base. 

The lack of any double-polarisation measurements from the beam-recoil group for this channel mean the mathematically "complete" constraint has not been achieved. This lack of previous data reflects the challenges in measurement of recoil nucleon polarisation, requiring a secondary rescattering of the ejectile nucleon in a spin-analysing polarimeter material. Such measurements are only feasible at high photon beam intensities\footnote{To achieve such measurements around four orders of magnitude larger statistics are required than for a typical beam-target measurements as only $\sim$2\% of the ejected events can be analysed with a practical thickness of nucleon scattering medium. 
}. The new $C_{x'}$ data presented here therefore provide an important cross check of convergence in the model predictions. Observables from the beam-recoil group give different sensitivities to the underlying reaction amplitudes, so even if determined with less precision than the beam-target measurements which dominate the database, they have the potential to provide important information on the road to convergence in the model predictions. Some constraint on systematics in both the reaction modelling and the fitted database can also be achieved as $C_{x'}$ is obtained with very different systematics to the beam-target double-polarisation data.\footnote{For example, correlated systematics could potentially arise as all the beam-target observables employed common methodologies in determining the degree of linear beam polarisation and MAMI/ELSA had common polarised target systems.} 



The photoproduction of $p\pi^0$ is the sister reaction to $n\pi^+$. For the $p\pi^0$ reaction the database is more complete. The differential cross-section, $\Sigma,P,T,G,H,C_{x'}$ observables have been determined over the full energy range of the new data and $E,O_{x'},O_{z'}$ for part of the range. Such $p\pi^0$ data are simultaneously fitted by theoretical models along with the $n\pi^+$ data. In combination,  sensitivities to the isospin of the contributing nucleon resonances and backgrounds can be achieved.

The leading models to interpret pion photoproduction data, and infer information on the nucleon resonance spectrum, are the SAID~\cite{SAID_MA19} and BnGa~\cite{BnGa1} frameworks, which are both based on partial wave analysis (PWA) methods fitted to the data. The latest iterations of these models (SAID MA19 and BnGa BG2019) experienced substantial change due to the inclusion of new data on the double-polarisation observable G, recently measured at CLAS~\cite{ZA19}.


 In this work we make a first determination of the $C_{x'}$ observable for the  $p(\gamma^{\odot},\vec{n}\pi^+)$ reaction. This is the first measurement of any observable from the beam-recoil double polarisation group for this reaction and the results are compared to nearly converged predictions from the leading partial wave analysis based theoretical models. The $C_{x'}$ observable is extracted using a bootstrap statistical technique which gives access to  statistical and systematic errors in a robust way. We compare the new data to the SAID~\cite{SAID_MA19} and BnGa~\cite{BnGa1} models, which are fitted to a database currently unconstrained by measurements of beam-recoil observables in this reaction channel.



\section{\label{sec:app}Experimental details}

\begin{figure}[!h]
\begin{center}
\includegraphics[width=0.48\textwidth,angle=0]{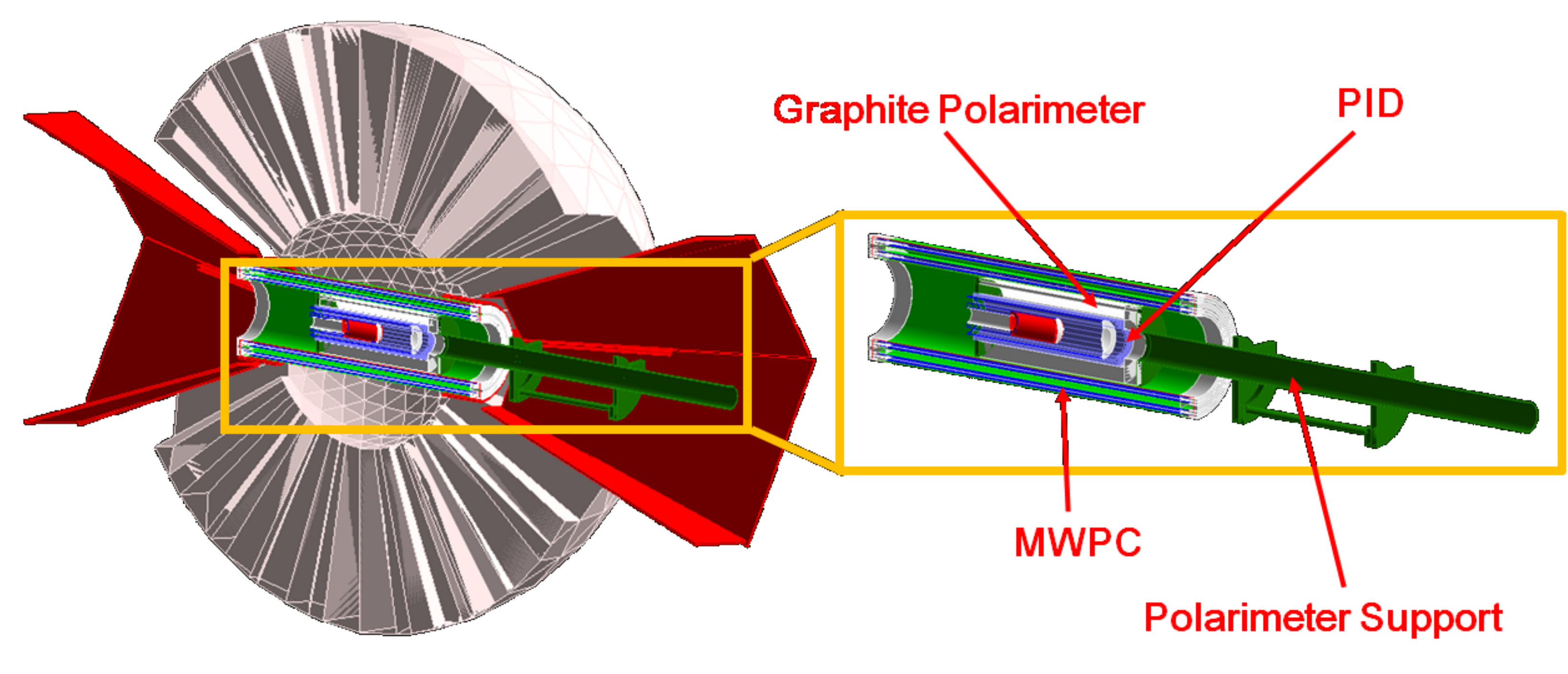}

\end{center}
\caption{Crystal Ball setup during the polarimeter beamtime. The cryogenic target (red cell) is surrounded by the PID barrel (blue), the graphite polarimeter (grey), the MWPC (blue/green) and the Crystal Ball (white).}
\label{Setup}
\end{figure}

The measurement employed a new, large acceptance, neutron polarimeter~\cite{proposal} within the Crystal Ball detector at the A2@MAMI~\cite{MAMI} facility during a 300 hour beamtime. A 1557~MeV longitudinally polarised electron beam impinged on either a thin amorphous (cobalt-iron alloy) or crystalline (diamond) radiator, producing circularly (alloy) or elliptically (diamond) polarised bremsstrahlung photons. The electron helicity was regularly flipped to produce a photon beam with equal amounts of both circular photon polarisations. As linear photon beam polarisation is not used to extract $C_{x'}^{n}$, equal flux from the two elliptical polarisation settings were combined to increase the circularly polarised photon yield\footnote{
Separate treatment of the events with two elliptical photon polarisations and with pure circular photon polarisation, gave consistent results within the achievable statistical accuracy.}.
The photons were energy-tagged ($\Delta E\sim~2$~MeV) by the Glasgow-Mainz Tagger~\cite{Tagg} and impinged on a 10~cm long liquid deuterium target cell.  Reaction products were detected by the Crystal Ball (CB)~\cite{CB}, a highly segmented NaI(Tl) photon calorimeter covering nearly 96\% of $4\pi$ steradians. For this experiment, a new bespoke 24 element, 7~cm diameter and  30~cm long plastic scintillator barrel (PID-POL)~\cite{PID} surrounded the target, with a smaller diameter than the earlier PID detector~\cite{PID}, but which provided similar particle identification capabilities. A 2.6~cm thick cylinder of analysing material (graphite) for nucleon polarimetry was placed around PID-POL, covering polar angles $12^{\circ} < \Theta < 150^{\circ}$ and occupying the space between PID-POL and the Multi Wire Proportional Chamber (MWPC)~\cite{MWPC}. The MWPC provided charged particle tracking for particles passing out of the graphite into the CB. At forward angles, an additional 2.6~cm thick graphite disk covered the range $2< \Theta < 12^{\circ}$~\cite{PID, MBPy, MBCx}. A GEANT4 visualisation of the experimental setup can be seen in Fig.~\ref{Setup}.

The cryogenic deuterium target provided a source of weakly bound protons and neutrons. The $d(\gamma,\pi^+\vec{n})n_{spec}$ events of interest consist of a primary charged pion track and a reconstructed neutron, which undergoes a $(n,p)$ charge-exchange reaction in the graphite to produce a secondary proton; the spectator neutron is not detected. The secondary proton gives signals in the MWPC and CB. The primary $\pi^{+}$ was identified using the correlation between the energy deposits in the PID and CB using $\Delta E-E$ analysis~\cite{PID} along with an associated charged track in the MWPC. The intercept of the primary $\pi^{+}$ track with the photon beamline allowed determination of the production vertex, and hence permitted the yield originating from the target cell windows to be removed. Neutron $^{12}$C$(n,p)$ charge exchange candidates required an absence of a PID-POL signal on the reconstructed neutron path, while having an associated track in the MWPC and signal in the CB from the scattered secondary proton. The incident neutron angle ($\Theta_{n}$) was determined from reaction kinematics using $E_{\gamma}$ and the production vertex coordinates. A distance of closest approach condition was imposed to ensure a crossing of the (reconstructed) neutron track and the secondary proton candidate track (measured with MWPC and CB). Once candidate pion and neutron tracks were identified, a kinematic fit was employed to increase the purity of the data sample and improve the determination of the reaction kinematics (see Ref ~\cite{mbMainz} for details).

\section{Determination of $C_{x'}^n$}
\begin{figure}[!h]
\begin{center}
\includegraphics[width=0.48\textwidth,angle=0]{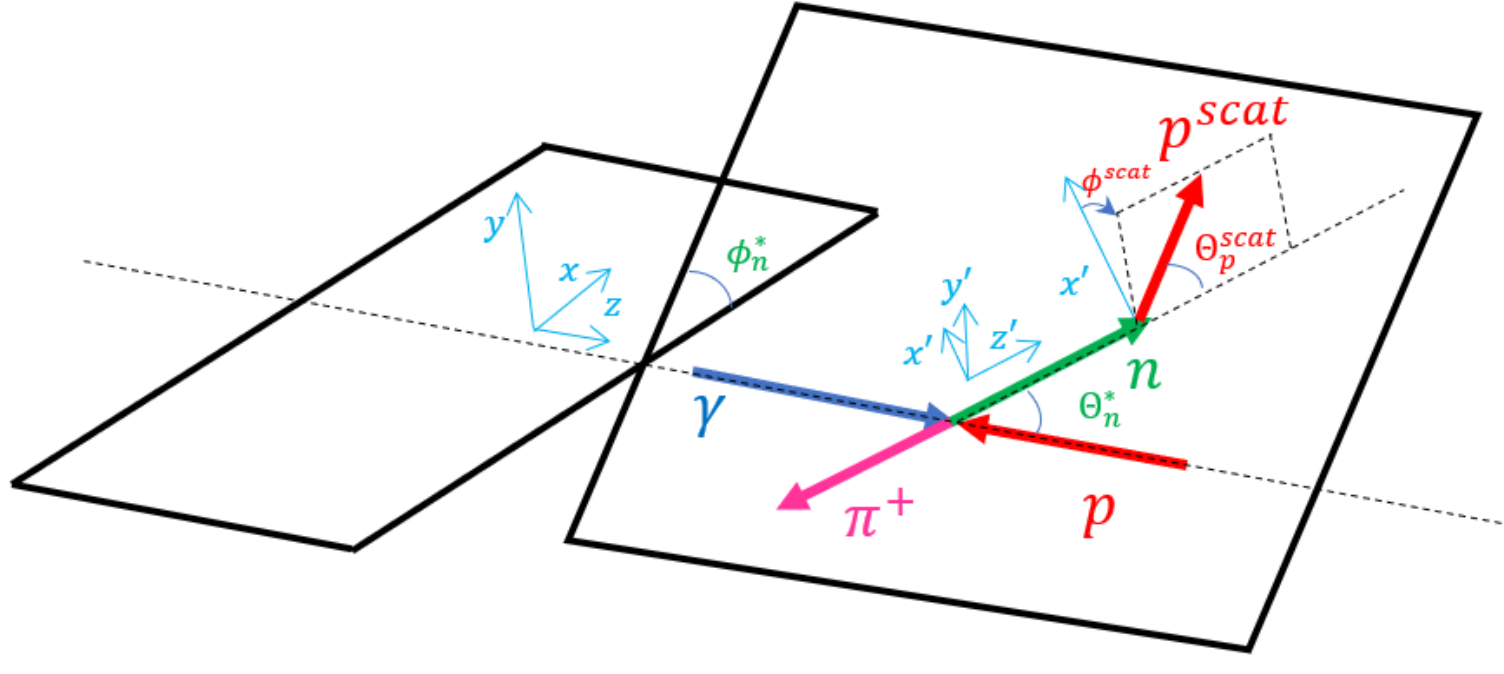}

\end{center}
\caption{Kinematics of the reaction in the centre-of-mass system. The z-axis is oriented along the photon beam, the y-axis is vertically upwards in the laboratory; the $z'$-axis is oriented along the ejectile neutron direction, and the $y'$-axis is perpendicular to the reaction plane.}
\label{Kin}
\end{figure}

The transferred neutron polarisation ($C_{x'}^n$) was determined through analysis of the neutron-spin dependent $^{12}$C$(n,p)$ reactions occurring in the graphite polarimeter. The spin-orbit component of the nucleon-nucleon interaction results in a $\phi$-anisotropy in the produced yield of secondary protons (see Ref.~\cite{SAID} for details). We followed the same procedure to determine $C_{x'}^n$, as described in Ref.~\cite{MBCx}, where the same observable was extracted for the deuteron photodisintegration reaction. The  $C_{x'}^n$ extraction employs a combination of log-likelihood and bootstrap techniques, as discussed below. The likelihood was defined as
\begin{equation}
    L_i=c_i\left[ 1+C^n_{x'}\cdot P_{\gamma,i}^{\odot}\cdot A_{y,i} \cdot \sin(\phi^{scat}_i)\right] A,
\end{equation}
where $c_i$ is a normalisation coefficient, $A$ is the detector acceptance, $C^n_{x'}$ is the spin transfer observable of interest, $P_{\gamma,i}^{\odot}$ is the degree of photon circular polarisation, $A_{y,i}$ is the neutron analysing power\footnote{$P_{\gamma,i}^{\odot}$ is photon energy dependent and $A_{y}$ depends on the ejected neutron energy and scattered proton polar angle. Both variables are evaluated on an event-by-event basis.} and $\phi^{scat}$ is the $\phi$ scattering angle in the primed frame, see Fig.~\ref{Kin}.
The log-likelihood function that was maximised to obtain the observable of interest is given by
\begin{equation}
    \log L=b+\sum_i \log\left[ 1 +C^n_{x'}\cdot P_{\gamma,i}^{\odot}\cdot A_{y,i} \cdot \sin(\phi^{scat}_i)  \right] \label{Eq:LogLike},
\end{equation}
where the constant $b$ is an observable-independent constant, which absorbs the normalisation coefficient and detector acceptance. The summation $(i)$ runs over all events. To reduce systematic dependencies, the events were only retained if $A_y(np)$ was above 0.1, and the proton scattering angle was in the range $\Theta^{scat}_{p} \in15~-~45^{\circ}$, where $\Theta^{scat}_p$ is the polar angle of the scattered proton relative to the direction of the neutron. The above cuts restrict the kinetic energy of the neutrons to be above 200~MeV, as well as their polar angle, $\Theta_{n}$ , to be larger than $\sim 40^{\circ}$, allowing them to be well within the acceptance of the polarimeter. Events meeting the conditions required a photon energy larger than 800~MeV and neutron centre-of-mass angle close to $90^{\circ}$. 

The extraction of $C_{x'}^n$ is simpler than the previous extraction of $P_y$ due to cancellations in the acceptance.\footnote{The determination of $P_{y}$ is obtained from a ($\cos(\phi^{scat}_i)$ dependence, which is sensitive to any systematics between forward and backward angle scattering~\cite{MBPy}. In contrast,  $C_{x'}^n$ has a ($\sin(\phi^{scat}_i)$ dependence (left-right asymmetry) for which the cylindrically symmetrical A2 setup has minimal acceptance/efficiency effects.} A regular ($\sim 1~s$) flipping of the photon helicity allowed us to reduce systematic effects associated with a temporal variation of the detector acceptance to negligible values. In addition, in the  determination of the $C_{x'}^n$ observable, the acceptance effects completely factor-out in the likelihood extraction (unlike the extraction of $P_y$). The use of an unbinned (in $\phi$ and $E_{\gamma}$) likelihood method eliminates bin-size related systematics arising from fits of trigonometric functions to binned $\phi^{scat}$-asymmetries. 

\begin{figure}[!h]
\begin{center}
\includegraphics[width=0.48\textwidth,angle=0]{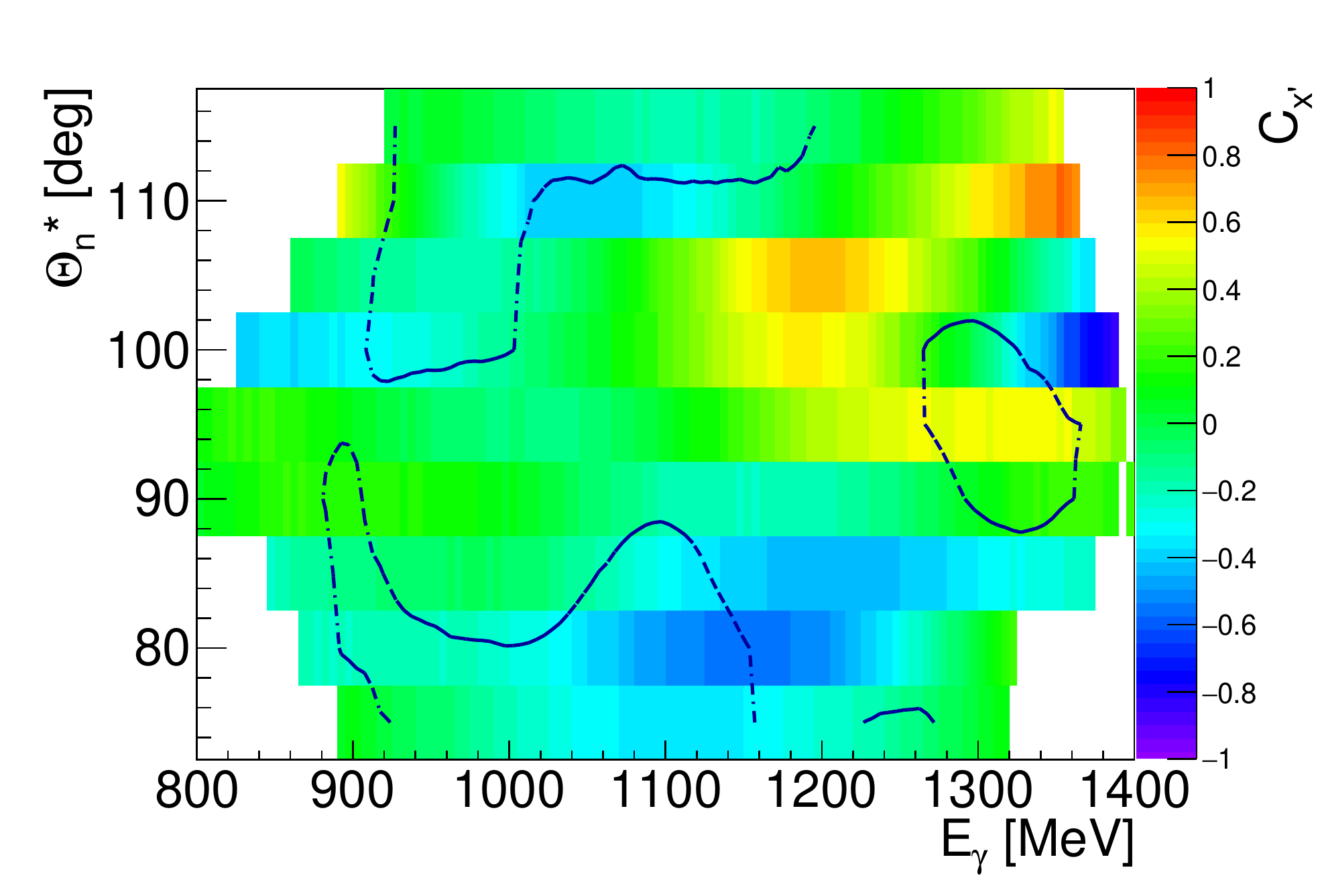}
\includegraphics[width=0.48\textwidth,angle=0]{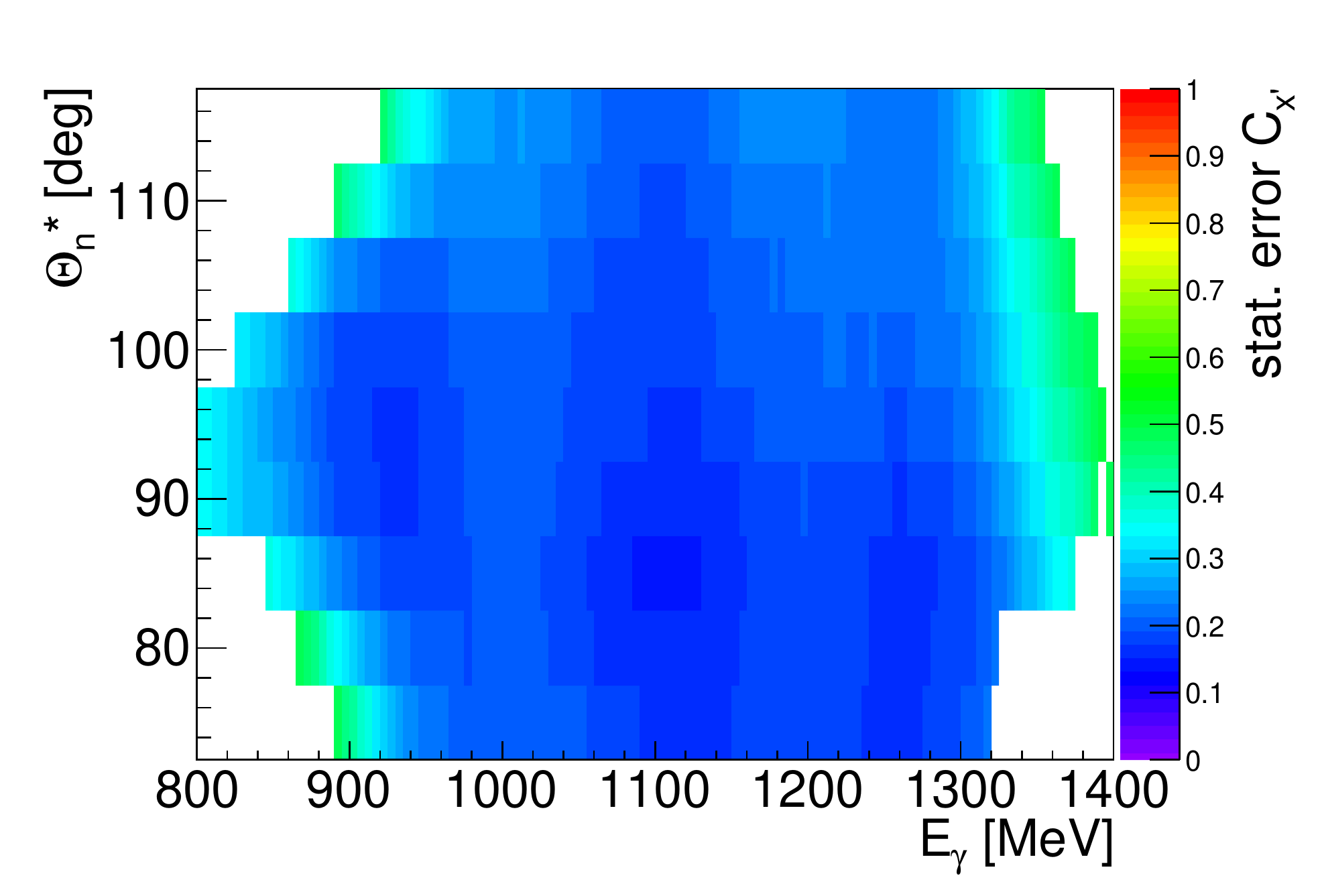}
\includegraphics[width=0.48\textwidth,angle=0]{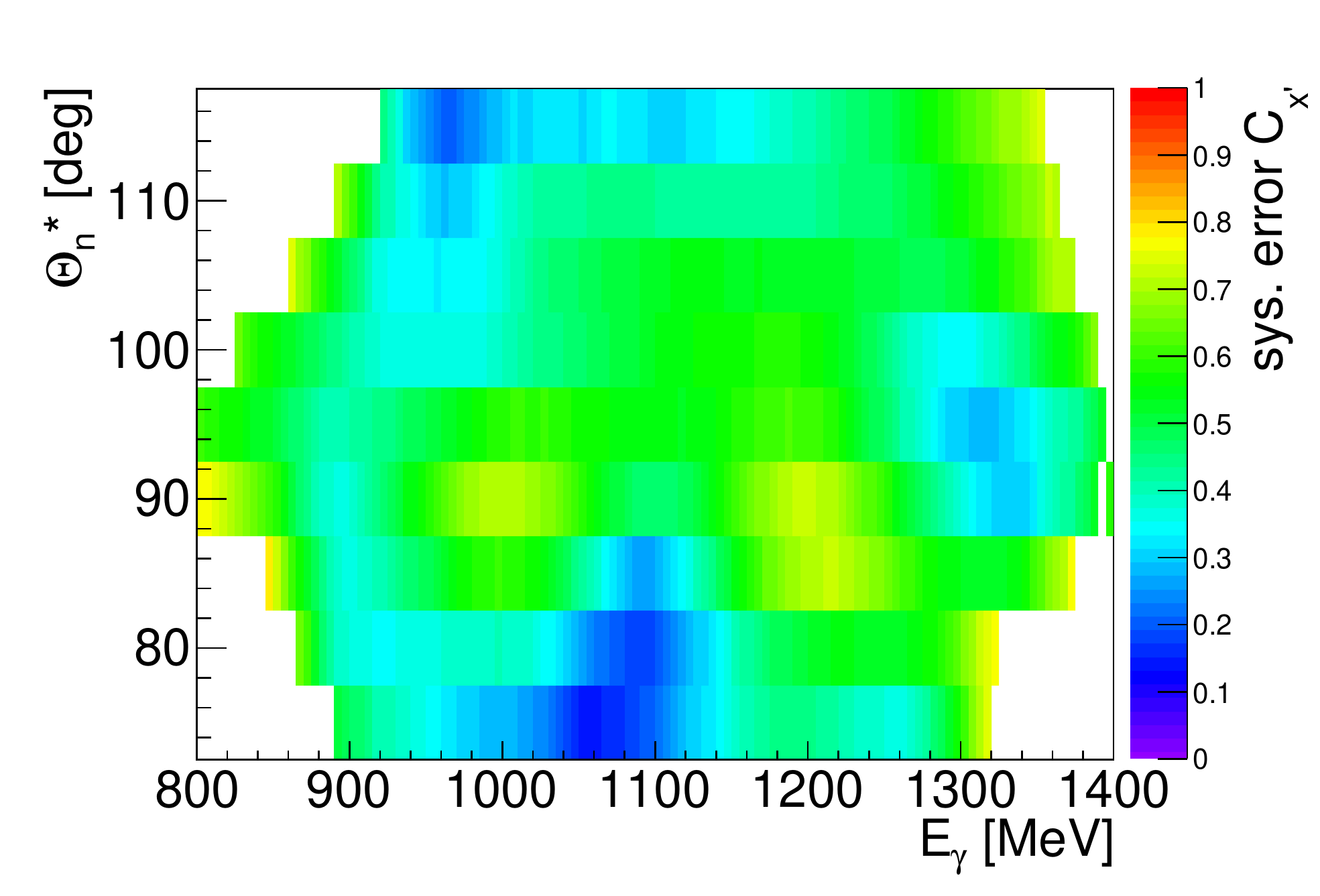}
\end{center}
\caption{Two-dimensional $C_{x'}^n$ dependence as a function of neutron centre-of-mass angle,$\Theta^{*}_{n}$ and photon energy $E_{\gamma}$~(top), statistical~(middle) and systematical~(bottom) uncertainties for this distribution. Systematic uncertainty contour lines at 0.4 are also shown on the top plot as dash-dot lines.}
\label{Cx2D}
\end{figure}

The spin transfer observable, $C_{x'}^n=f(\Theta_N^*,E_{\gamma})$, was determined in $10^{\circ}$ neutron centre-of-mass (CM) angle bins using a likelihood-extraction in which a smooth energy dependence is assumed. To get robust error evaluations within the likelihood method we employed a bootstrap procedure~\cite{Ale}. This method involves randomly selecting $N$ events out of our sample of $N$ events allowing repetitions\footnote{For example, in the case of the $80^{\circ}$ bin it is 9200 random events out of a 9200 event sample.}. For each combination, a likelihood fit is carried out to extract $C_{x'}^n$. Each fit produces a smooth function $C_{x'}^n=f(E_{\gamma})$ for the given $\Theta_n$ bin. Multiple repetitions of this procedure enable determination of the most likely $C_{x'}$ and its associated errors $C_{x'}^n$. These are presented in Fig.~\ref{Cx2D} (middle). Additional studies of systematic effects were obtained by relaxing the analysis cuts. The systematic errors are extracted from the resulting variations in the extracted values of $C_{x'}$. The magnitude of the systematic errors in each bin are influenced by the achievable statistical accuracy in the bin. The extracted systematic errors are also shown in Fig.~\ref{Cx2D} (bottom)\footnote{The dataset corresponds to kinematical regions where $C_{x'}^n$ statistical errors are less than 0.5 or systematical errors are smaller than 0.8.}.

\section{\label{sec:sigma} Results}

Our $C_{x'}^n$ results, along with associated statistical and estimated systematic errors, are presented in Fig.~\ref{Cx2D} as a function of photon energy and neutron CM polar angle. The new data cover photon energy bins from 800~MeV up to 1400~MeV and CM angles for the neutron of 70-120$^{\circ}$. The magnitude of $C_{x'}^n$ is seen to be small over much of the sampled phase space, albeit with indications of localised regions of both high and low $C_{x'}^n$. The statistical and systematic uncertainties are both largely uniform over much of the phase space, but with expected deterioration close to the edge of the polarimeter acceptance. 

\begin{figure}[!h]
\begin{center}
\includegraphics[width=0.5\textwidth,angle=0]{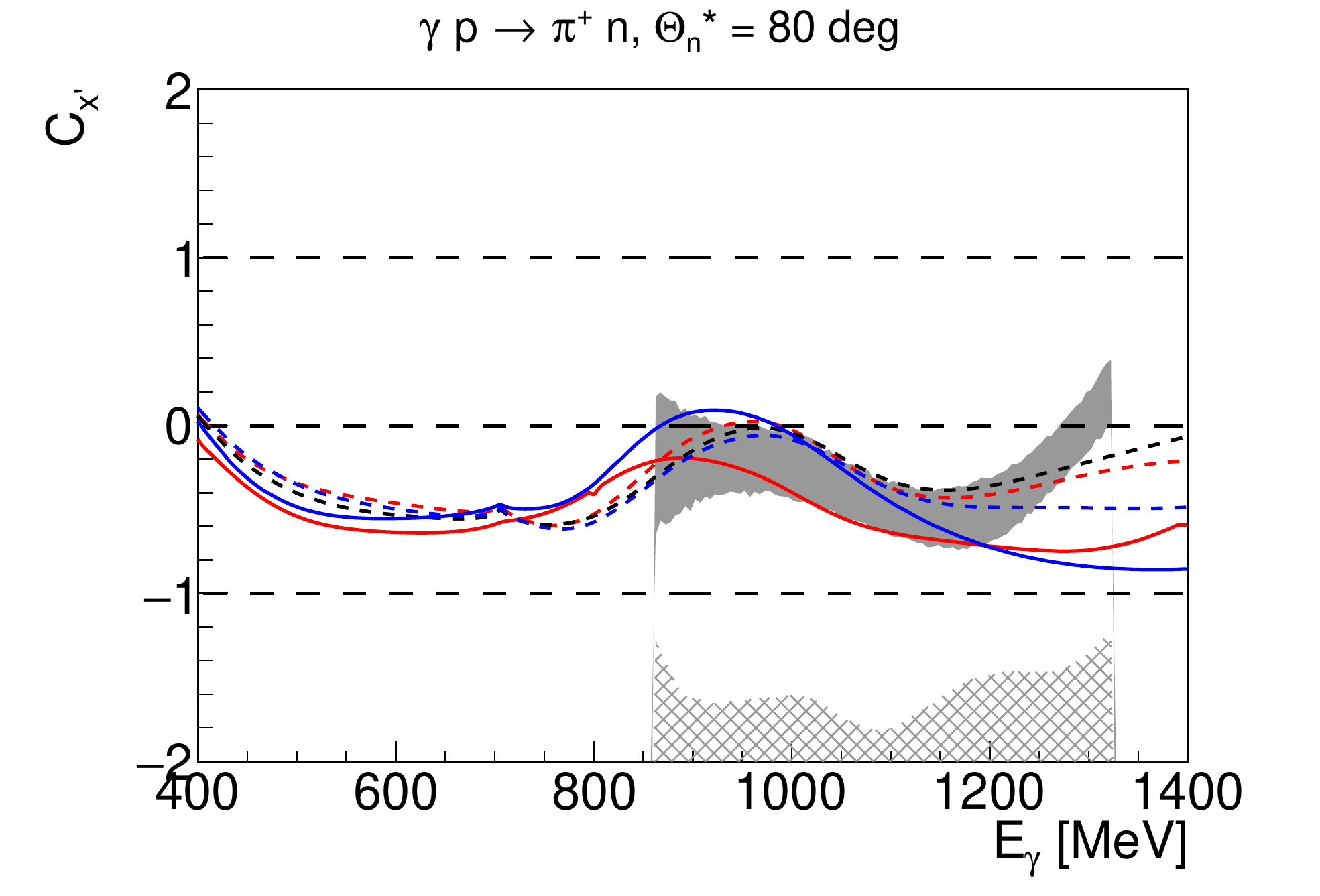}
\end{center}
\caption{The neutron spin transfer observable $C_{x'}^n$ for the $75^{\circ}-85^{\circ}$, neutron CM angles as a function of photon beam energy. The dashed area represents the data together with its statistical uncertainty, while the hatched area shows the systematic uncertainty (the data and uncertainties are a 1D slice of Fig.~\ref{Cx2D}).The solid lines are the SAID PWA solutions, SAID-2019(red)~\cite{SAID_MA19}, SAID-2012(blue) solution~\cite{SAID_SM12}; the dashed lines are the BnGa solutions, BnGa-19(red)~\cite{ZA19}, BnGa-2014-1(blue), BnGa-2011-1(black).}
\label{Cx2}
\end{figure}

\begin{figure}[!h]
\begin{center}
\includegraphics[width=0.5\textwidth,angle=0]{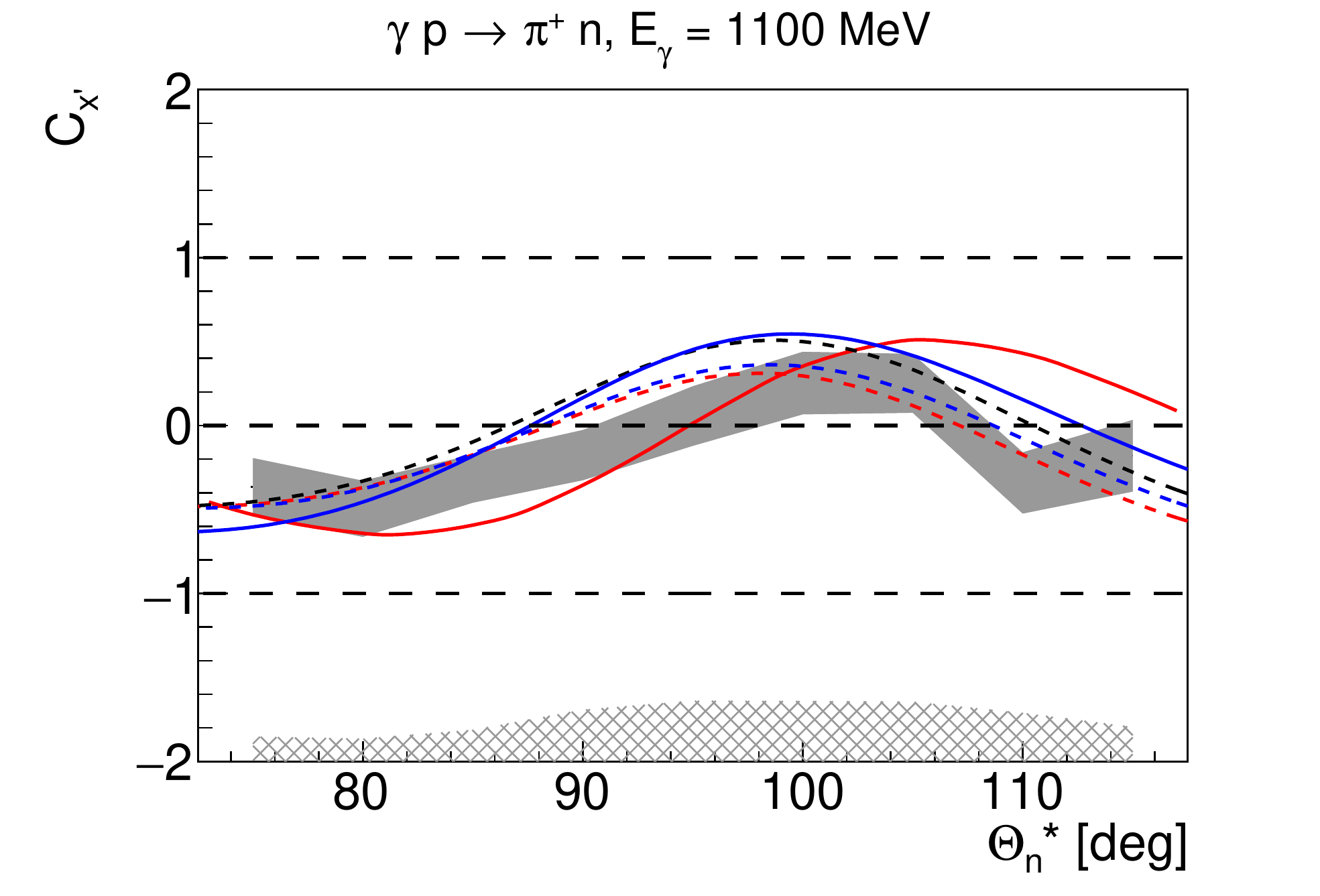}
\includegraphics[width=0.5\textwidth,angle=0]{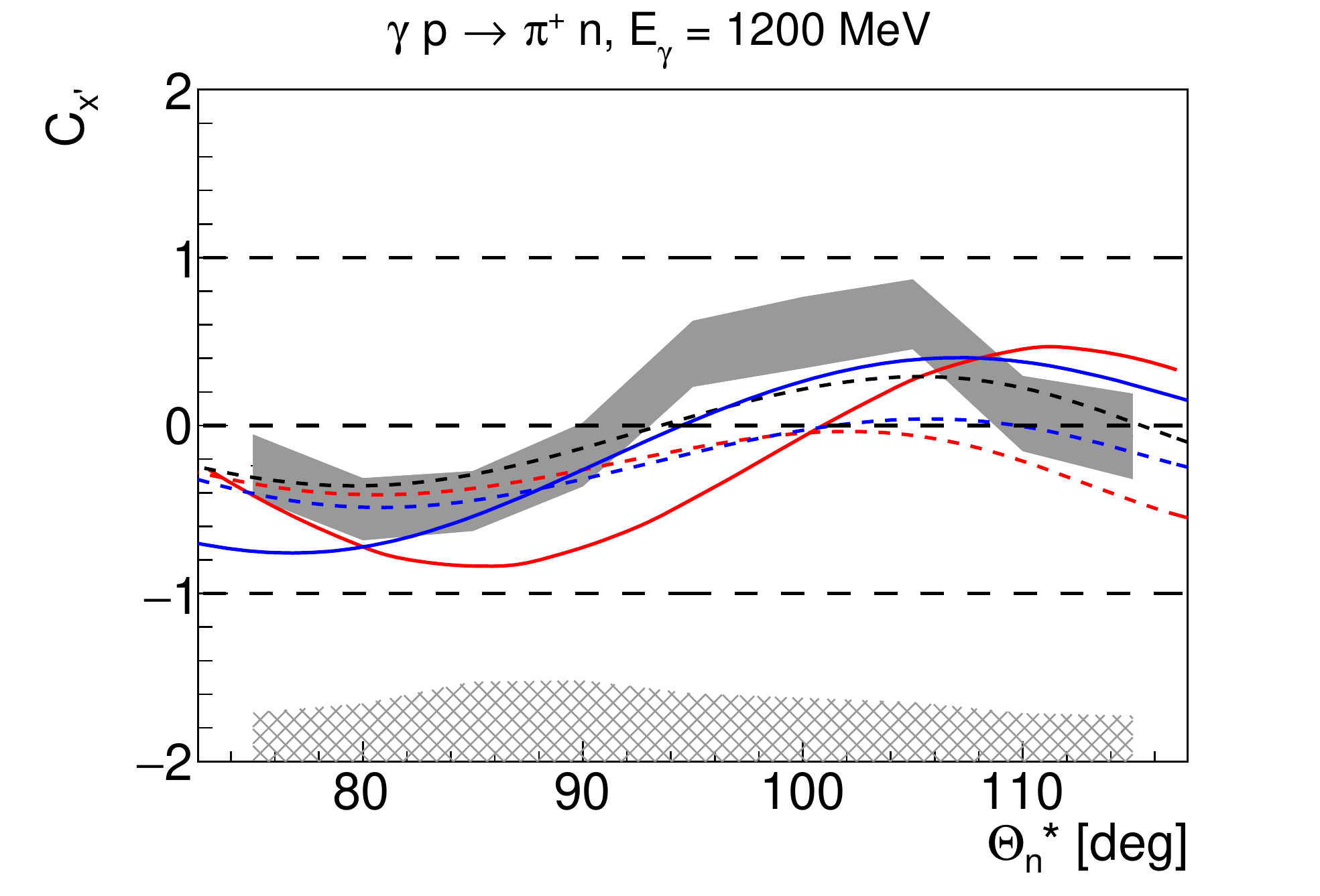}
\end{center}
\caption{The neutron spin transfer observable $C_{x'}^n$ for the 1100~MeV~(top) and 1200~MeV~(bottom) beam photon energies as a function of neutron CM angle ($5^{\circ}$ bin spacing). The dashed area represents the data together with its statistical uncertainty, while the hatched area shows the systematic uncertainty. The theory lines are the same as in Fig.~\ref{Cx2}.}
\label{Cx3}
\end{figure}

Example comparisons of the new data with recent solutions of the  SAID~\cite{SAID_MA19} and BnGa~\cite{ZA19} partial wave fits are shown in Fig.~\ref{Cx2} and \ref{Cx3}. 
In Fig.~\ref{Cx2} the data are presented as a function of photon energy for a fixed angular bin of 80$^{\circ}$. For photon energies up to 1200~MeV the data are well described by all SAID and BnGa solutions (see figure caption) within the achievable experimental errors\footnote{The structure seen in theory curves around $E_{\gamma}\sim 700$~MeV originates from the $\eta$-meson production threshold.}. Above this, the data may suggest higher $C_{x'}^n$ than both model predictions - albeit in regions where the estimated systematic uncertainties are large.

The angular distribution of $C_{x'}^n$ at 1100 MeV is presented in Fig. \ref{Cx3} (top), compared to the model predictions. The observed zero crossing at around $90^{\circ}$ is broadly  consistent with all model predictions, however the second crossing around $108^{\circ}$ is in better agreement with the latest BnGa solution. The angular distribution of $C_{x'}^n$ at 1200 MeV, Fig. \ref{Cx3} (bottom), shows  a broadly similar shape to the lower energy bin, but with indications of regions of higher positive $C_{x'}^n$. 
The divergence between the models is larger for this higher photon energy bin. The most recent SAID and BnGa fits (red solid and red-dashed respectively) include the large new {\it G-} dataset~\cite{ZA19} for $n\pi^{+}$, but it's inclusion results in both models giving a poorer description of this independent $C_{x'}^n$ variable. This suggests constraint from a different double-polarisation group, even with poorer coverage and accuracy, has the potential to provide new information. Future fits including the new data may improve convergence between the models for this photon energy region. 
 
The statistical and systematic  uncertainties in $C_{x'}^n$ extraction could be significantly reduced in the future due to recent upgrades in the MAMI beam intensity. Accurate beam-recoil data appears important even when, as is the case here, an extensive database of beam-target double-polarisation observables has already been obtained.



\section{\label{sec:final} Summary}

This work provides the first  measurement of a beam-recoil double-polarisation observable, $C_{x'}^n$, for $\pi^+$ photoproduction off the proton, obtained for photon energies 800-1300 MeV. The measurement provides an important check of the leading partial wave analysis based models which fit the world data on photo-meson production from the nucleon to extract the nucleon excitation spectrum. The general trends of the new $C_{x'}^n$ data are reproduced by the two leading model predictions (SAID and BnGa), although deviations between the predictions remain in this photon energy region. The inclusion of the most recent beam-target double- polarisation data in the fitted database, resulted in a poorer description of the new $C_{x'}^n$ data. This suggests beam-recoil observables give sensitivities to reaction amplitudes that are not currently well constrained by world data and are an important addition to the database on the path towards model convergence in the extracted amplitudes.


\section{Acknowledgement}

In accordance with UK data management policy, all data are available for downloading from PURE~\cite{PURE}. This work has been supported by the U.K. STFC (ST/V002570/1, ST/L00478X/2, ST/V001035/1, ST/P004385/2, ST/T002077/1, ST/L005824/1, 57071/1, 50727/1 ) grants, the Deutsche Forschungsgemeinschaft (SFB443, SFB/TR16, and SFB1044), DFG-RFBR (Grant No. 09-02-91330), Schweizerischer Nationalfonds (Contracts No. 200020-175807, No. 200020-156983, No. 132799, No. 121781, No. 117601), the U.S. Department of Energy (Offices of Science and Nuclear Physics, Awards No. DE-SC0014323, DEFG02-99-ER41110, No. DE-FG02-88ER40415, No. DEFG02-01-ER41194) and National Science Foundation (Grants NSF OISE-1358175; PHY-1039130, PHY-1714833, No. IIA-1358175, PHY-2012940), INFN (Italy), and NSERC of Canada (Grant No. FRN-SAPPJ2015-00023). This infrastructure is part of a project that has received funding from the European Union’s Horizon 2020 research and innovation programme under grant agreement No 824093.


\end{document}